\documentclass[times]{jaa}
\usepackage{txfonts}
\usepackage{graphicx}
\newcommand{\newblock}{}
\usepackage[authoryear]{natbib}
\begin{document}
\title[Variability of OI 090.4]{Variability of OI 090.4}
\author[S.M. Hu et al.]%
       {ShaoMing Hu\thanks{e-mail:husm@sdu.edu.cn}, Xu Chen and DiFu Guo\\
       School of Space Science and Physics, Shandong University, Weihai\\ 180 Cultural West Road, Weihai, Shandong 264209, China}
\maketitle
\label{firstpage}
\begin{abstract}
OI 090.4 was monitored on 21 nights from 2006 to 2012 for studying the variability. Strong variations occurred during the past 6 years. The long-term variability amplitude is consistent with previous results. Microvariability was analyzed for 43 intra-night light curves. 30 out of 43 light curves showed microvariability by C and F tests analysis.
\end{abstract}
\section{Introduction}

Blazars are a subclass of Active Galactic Nuclei (AGN), known to be variable on different time-scales across
the whole electromagnetic spectrum \citep{ulrich97}.
It consists of flat spectrum radio
quasars (FSRQs) and BL Lacertae objects (BL Lacs).
Until now, except for the Mpc scale
radio-jets and lobes, the other components within the canonical model remain spatially unresolved
to astronomical observations in the vast majority of AGNs, but variability study could
constrain the size and physical mechanism of the emitting
regions. So variability becomes a powerful tool to study those innermost
regions.

OI 090.4 (PKS 0754+100) was identified as a BL Lac object by \cite{Tapia77}. \cite{ghosh95} suggested it is a Low frequency-peaked BL Lac object (LBL) by its multifrequency spectrum.
This is another BL Lac whose redshift is still uncertain
\citep{fu2000}. \cite{baumert} reported a $\sim2$ mag optical variations over long time scales. \cite{fan} reported a $\sim3$ mag variability in 10 yr, with smaller
variations up to 1 mag in about 1 year. Other authors reported its variability on different time-scales (e.g. Pica et al.1988; Smith et al. 1987; Ghosh et al. 2000; Xie et al. 2004). But most observations was sparse for microvariability study. \cite{cellone} investigated its inter-night and intra-night variability (microvariability) by monitoring on 5 nights. It varied 0.36mag and 0.42mag in R and V band, respectively. Microvariability was detected on 4 out of 5 nights. In order to investigate the properties of variability, we carried out the monitoring of OI 090.4 from 2006 to 2012. In section 2, the observations and data reduction was given, while in section 3 we
showed the results and discussion.

\section{Observations and data reduction}

OI 090.4 was monitored from 2006 Feb. to 2012 Feb. using 1.0m telescope at Weihai observatory of Shandong University, 2.16m and 80cm telescopes at Xinglong station of NAOC. In order to study the microvariability and color behavior of the source, we observed it repeated with Johnson/Cousins filters V, R and I sequently as long as possible within one night, and we did observations as quickly as possible for high temporal resolution. For obtaining enough signal to noise ratio, the exposure time was set between 80s and 480s depending on the size of telescope, filters and weather conditions. 886 images were obtained on 21 nights. The images were processed automatically using an Interactive Data Language (IDL) procedure developed by ourself based on the NASA IDL astronomical libraries. All images were reduced by bias, flat correction and aperture photometry. The aperture size was set to 6.3$''$ for images from Weihai and Xinglong 80cm telescope, and 8$''$ was set for images from 2.16m telescope because of relative larger seeing. Stars A and B \citep{fiorucci} were chosen for comparison to do differential photometry.

\section{Results and discussion}

\begin{figure}[b]
    \begin{tabular}{lr}
    \begin{minipage}[b]{6cm}
    \includegraphics[width=4.9cm]{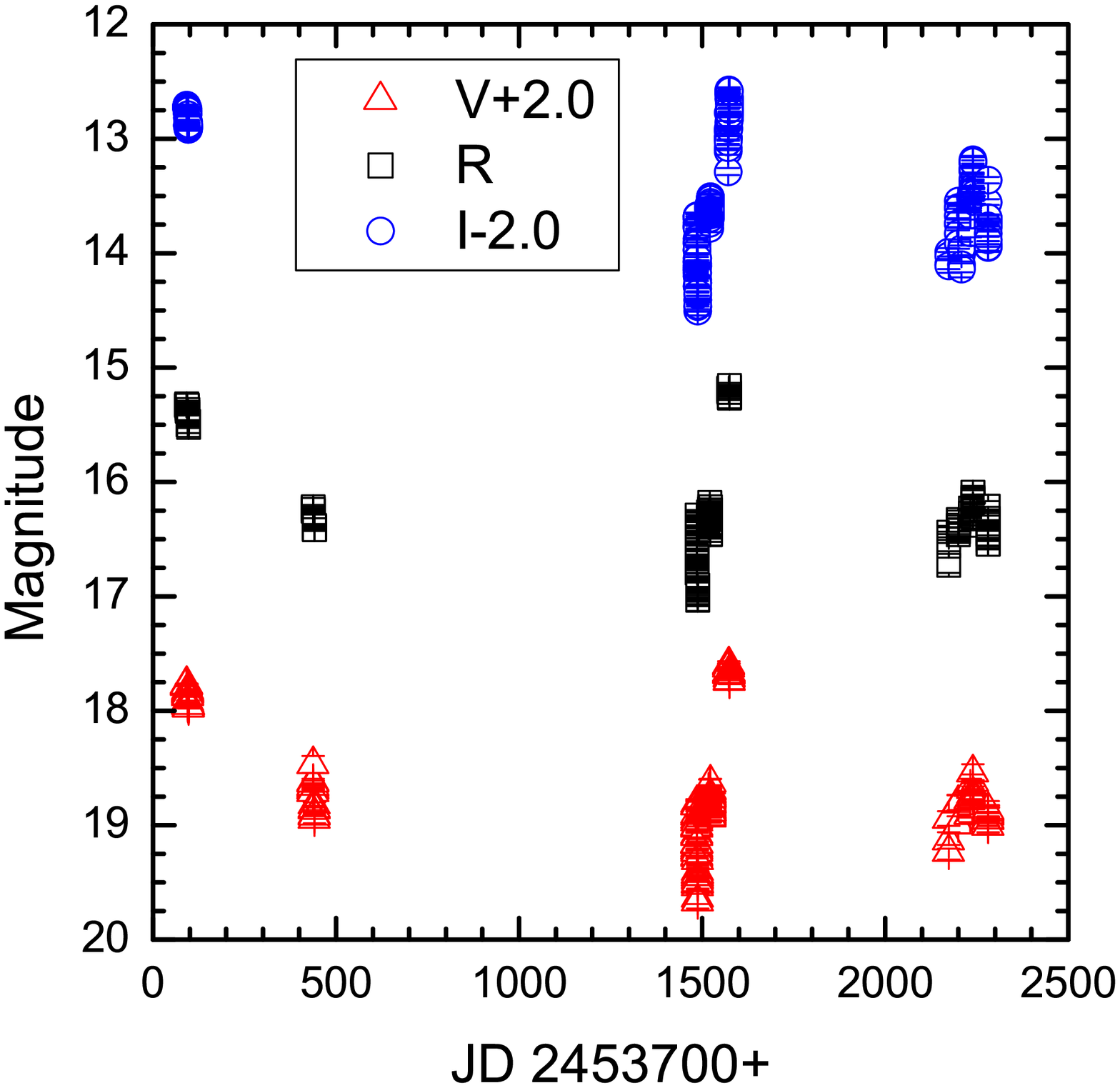}
    \caption{Light curves of OI 090.4 from 2006 Feb. to 2012 Feb.}
    \label{fig1}
    \end{minipage}

    \begin{minipage}[b]{6cm}
    \includegraphics[width=4.9cm]{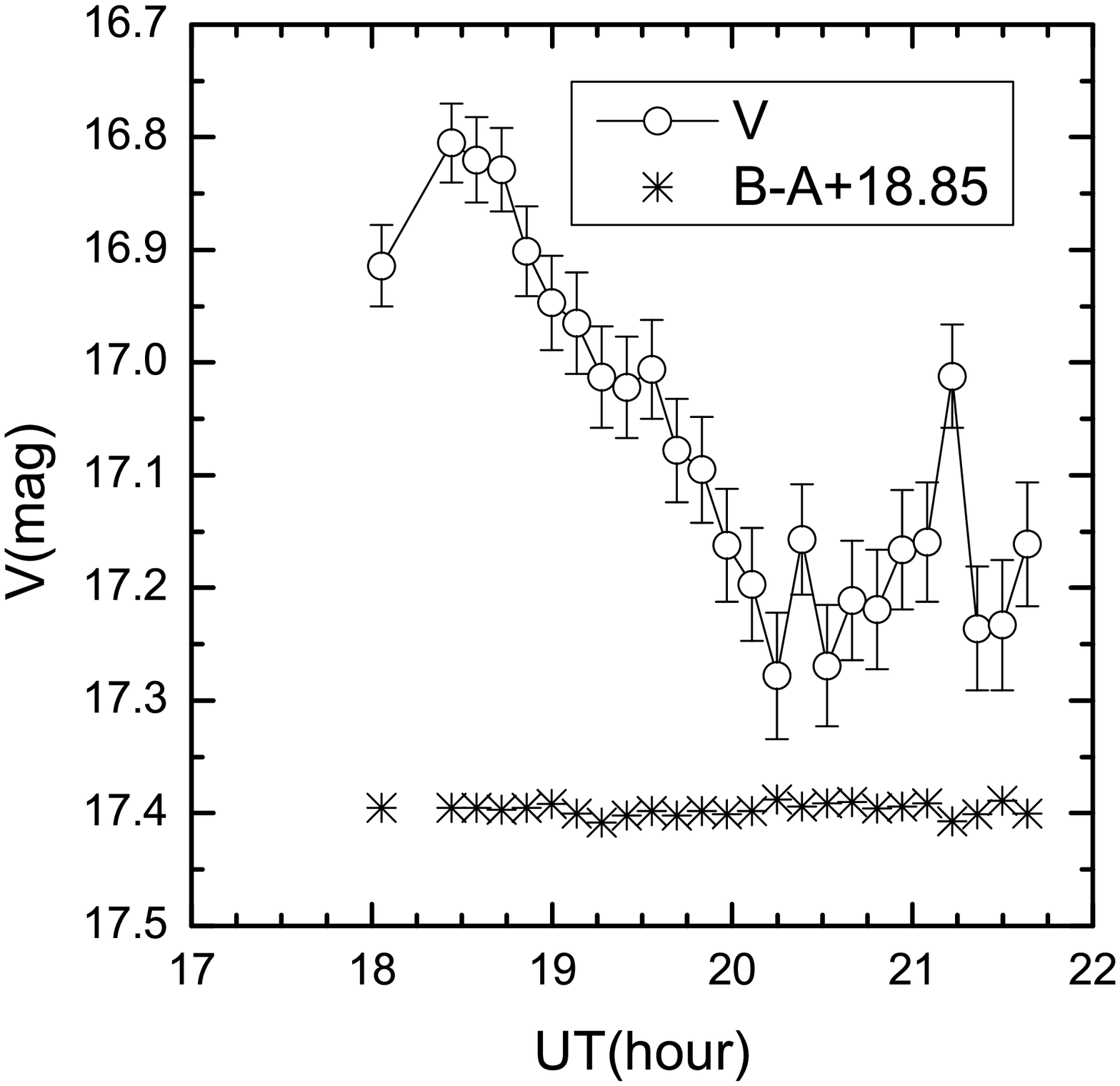}
    \caption{Light curve in V band on 2009 Dec. 20}
    \label{fig2}
    \end{minipage}
\end{tabular}
\end{figure}

The data were filtered for analysis accuracy. Firstly, we discarded the data whose photometry error is larger than 0.08mag. Then we discarded the data whose related check star magnitude is out of two times of their standard deviation within the whole night. 746 observations in V, R and I bands were remained. Long-term light curves were shown in Fig.\ref{fig1}. Red triangles, black boxes and blue dots
indicate the variations in V, R and I band, respectively. Strong variability occurred during the past 6 years. It changed 2.07mag (from 15.59 to 17.66), 1.88mag (from 15.15 to 17.03), and 1.93mag (from 14.57 to 16.50) in V, R and I band, respectively. These variability amplitude is consistent with the results reported by \cite{baumert}, \cite{noble} (2.67mag in V band) and \cite{fan}.

Small variations superposed on the long term variability. Our intense observations enabled us to study the microvariability. C test \citep{romero} and F test \citep{diego10} were used to detect the microvariability for light curves, whose data points were more than 8 within one night. We will claim that microvariability was detected (not detected) if it was judged variable (not variable) by both tests (taken $99\%$ as the confidence level). Fig.\ref{fig2} shows an example of microvariability light curve in V band on 2009 Dec. 20. Variability amplitude was calculated for light curves with microvariability \citep{heidt96}. All results were listed in table\ref{table1}. Column 1 denotes the observation date, and band (Col.2), number of observations (Col.3), duration of light curve (Col.4), C value (Col.5), F test value (Col.6), critical F value corresponding to $99\%$ confidence level (Col.7), label of microvariability (Col.8) and variability amplitude (Col.9) are presented one by one.  30 light curves out of 43 have microvariability, and the duty cycle is about $70\%$. Our result supports the result of \cite{cellone} ($80\%$).

Physical causes of blazar variability are still under debate. But intrinsic models(e.g. hotpots models, shock-in-jet models and geometrical models) could widely explain optical variability, and we prefer shock-in-jet plus geometrical model. Many Investigations (Heidt \& wagner, 1996,1998; Gaur et
al., 2012) \nocite{heidt96,heidt98,Gaur2012b} showed that HBLs have less optical variability than that of LBLs. The scenario of stronger magnetic fields in HBLs \citep{romero} is a possible physical reason. More research in this field is crucial to study the mechanism of variability.

%\verb+\setcounter\table{+{\em 1}\verb+}
\begin{table*}
\scriptsize
\caption{Microvariability of OI 090.4 }
\label{table1}
  \centering
  \begin{tabular}{c c r r r r r c c}
\hline\hline
  Date & Band & N  & $\Delta T$ & C & F & $ F_{99} $ & Var?  & A \\
  \hline
2009-12-20 & I  &  27 &  3.78 &    5.742 &  32.974 &   2.554 &    Y      &   3.868\\
2009-12-20 & R  &  27 &  3.77 &   18.944 & 358.883 &   2.554 &    Y      &   3.059\\
2009-12-20 & V  &  24 &  3.58 &   31.681 &1003.717 &   2.719 &    Y      &   2.770\\
2009-12-21 & I  &  18 &  4.01 &    3.930 &  15.447 &   3.242 &    Y      &   2.803\\
2009-12-21 & R  &  17 &  4.02 &    1.212 &   1.469 &   3.372 &    N     &   \\
2009-12-21 & V  &  15 &  3.78 &    2.867 &   8.221 &   3.698 &    Y      &   2.033\\
2010-01-21 & I  &  11 &  1.39 &    1.679 &   2.820 &   4.849 &    N      &   \\
2010-01-21 & R  &  11 &  1.39 &    4.872 &  23.737 &   4.849 &    Y      &   0.835\\
2010-01-21 & V  &  11 &  1.39 &    6.301 &  39.697 &   4.849 &    Y      &   1.080\\
2010-01-23 & I  &  34 &  4.60 &    1.204 &   1.450 &   2.287 &    N     &   \\
2010-01-23 & R  &  34 &  4.60 &    6.902 &  47.631 &   2.287 &    Y      &   1.527\\
2010-01-23 & V  &  35 &  4.75 &    7.083 &  50.172 &   2.258 &    Y     &   1.667\\
2010-01-25 & I  &  39 &  5.66 &    2.482 &   6.160 &   2.157 &    ?      &   \\
2010-01-25 & R  &  39 &  5.91 &   10.485 & 109.936 &   2.157 &     Y     &   1.512\\
2010-01-25 & V  &  40 &  6.10 &   12.211 & 149.116 &   2.135 &     Y     &   1.826\\
2010-03-16 & I  &  20 &  4.51 &    2.410 &   5.808 &   3.027 &     ?   &   \\
2010-03-18 & I  &  14 &  3.54 &    0.712 &   0.507 &   3.905 &     N     &   \\
2010-03-18 & R  &   9 &  3.26 &    4.099 &  16.798 &   6.029 &     Y     &   0.749\\
2010-03-18 & V  &   9 &  2.99 &    2.143 &   4.594 &   6.029 &     N     &   \\
2011-12-03 & I  &   8 &  1.08 &    9.261 &  85.772 &   6.993 &     Y     &   2.372\\
2012-01-13 & I  &  12 &  3.17 &    2.139 &   4.575 &   4.462 &     ?     &   \\
2012-01-13 & R  &   9 &  2.46 &    4.733 &  22.398 &   6.029 &     Y     &   1.333\\
2012-02-23 & I  &  14 &  3.35 &    3.783 &  14.313 &   3.905 &     Y     &   3.707\\
2012-02-23 & R  &  15 &  3.35 &   16.108 & 259.464 &   3.698 &     Y     &   2.044\\
2012-02-23 & V  &  12 &  3.59 &    5.439 &  29.586 &   4.462 &     Y     &   0.818\\
2007-02-07 & V  &   8 &  2.11 &   20.501 & 420.274 &   6.993 &     Y     &   1.466\\
2007-02-09 & R  &   8 &  2.63 &    5.236 &  27.417 &   6.993 &     Y     &   0.267\\
2007-02-09 & V  &   8 &  3.01 &    7.592 &  57.635 &   6.993 &     Y     &   0.763\\
2006-02-25 & I  &   9 &  3.22 &    5.111 &  26.124 &   6.029 &     Y    &   0.223\\
2006-02-25 & R  &  10 &  3.24 &    2.412 &   5.817 &   5.351 &     ?     &   \\
2006-02-25 & V  &  10 &  3.29 &    2.693 &   7.252 &   5.351 &     Y     &   0.106\\
2006-02-26 & I  &   9 &  2.72 &    7.952 &  63.228 &   6.029 &     Y     &   0.985\\
2006-02-26 & R  &  10 &  2.74 &    2.435 &   5.928 &   5.351 &     ?     &   \\
2006-02-28 & I  &   9 &  2.67 &    4.070 &  16.566 &   6.029 &     Y    &   0.767\\
2006-02-28 & R  &   9 &  2.69 &    4.646 &  21.583 &   6.029 &     Y     &   0.271\\
2006-02-28 & V  &  10 &  2.75 &    4.880 &  23.812 &   5.351 &     Y     &   0.333\\
2006-03-01 & I  &  10 &  2.45 &    2.478 &   6.140 &   5.351 &     ?     &   \\
2006-03-01 & R  &   9 &  2.47 &    4.364 &  19.043 &   6.029 &     Y     &   0.199\\
2006-03-01 & V  &  10 &  2.47 &    1.725 &   2.974 &   5.351 &     N     &   \\
2006-03-02 & I  &  10 &  2.30 &    2.696 &   7.269 &   5.351 &     Y     &   0.417\\
2006-03-02 & R  &  10 &  2.31 &    5.481 &  30.042 &   5.351 &     Y     &   0.309\\
2006-03-02 & V  &   9 &  2.33 &    2.886 &   8.328 &   6.029 &     Y     &   0.185\\
\hline
\end{tabular}
\end{table*}

\noindent{{\textbf{Acknowledgements:}}} \\
\noindent{We wish to thank NAOC for time allocation to support this program. This work was supported by NSFC under grant Nos. 11143012, 11203016, 10778619 and 10778701, and by the NSF of Shandong Province under grant No. ZR2012AQ008.

\label{lastpage}

\end{document}